\documentclass{aa}
\usepackage{graphicx,natbib,amssymb}
\usepackage{txfonts}
\bibpunct[, ]{(}{)}{,}{a}{}{,}

   \title{\emph{XMM-Newton} observations of the BL~Lac MS\,0205.7+3509:
          a dense, low-metallicity absorber}

   \titlerunning{A low-metallicity absorber in the foreground of MS\,0205.7+3509}

   \author{D.~Watson\inst{1,2} \and
           B.~McBreen\inst{3} \and
           L.~Hanlon\inst{3} \and
           J.~N.~Reeves\inst{4,5} \and
           N.~Smith\inst{6} \and
           E.~Perlman\inst{7,8} \and
           J.~Stocke\inst{9} \and
           T.~A.~Rector\inst{10,11}
          }

   \offprints{D.~Watson, \texttt{darach@astro.ku.dk}}

   \institute{Niels Bohr Institute, Astronomical Observatory, University
              of Copenhagen, Juliane-Maries Vej 30, DK-2100 Copenhagen \O,
              Denmark
         \and
              X-Ray Astronomy Group, Department of Physics and Astronomy,
              Leicester University, Leicester LE1 7RH, UK
         \and
              Dept.\ of Experimental Physics, University College Dublin,
              Dublin 4, Ireland
         \and
              Laboratory for High Energy Astrophysics, Code 662,
              NASA Goddard Space Flight Center, Greenbelt, MD 20771, USA
         \and 
              Universities Space Research Association
         \and
              Dept.\ of Applied Physics and Instrumentation,
              Cork Institute of Technology, Cork, Ireland
         \and
              Joint Center for Astrophysics, University of Maryland, Baltimore County, 1000 Hilltop Circle, Baltimore, MD  21250, USA
         \and
              Johns Hopkins University, Baltimore, MD 21218, USA
         \and
              Center for Astrophysics and Space Astronomy, University of Colorado,
              Boulder, CO 80309-0389, USA
         \and
              National Optical Astronomy Observatory, 950 N. Cherry Ave.,\ Tucson, AZ 85719 USA
         \and
              University of Alaska, 3211 Providence Dr., BMB 212, Anchorage, AK 99508, USA
}

   \date{Received / Accepted }

   \abstract{The high-frequency-peaked BL~Lac,
	     MS\,0205.7+3509 was observed twice with
	     \emph{XMM-Newton}.  Both X-ray spectra are
             synchrotron-dominated, with mean 0.2--10\,keV fluxes of
	     $2.80\pm0.01$ and
	     $3.34\pm0.02\times10^{-12}$\,erg\,cm$^{-2}$\,s$^{-1}$. The
	     X-ray spectra are well fit by a power-law with absorption above
	     the Galactic value, however no absorption edges are detected,
	     implying a low metallicity absorber ($Z_{\sun} =
	     0.04^{+0.03}_{-0.01}$) or an absorber with redshift above one
	     (best-fit $z=2.1$ for an absorber with solar abundances).  In
	     either case the absorbing column density must be
	     $\sim9\times10^{21}$\,cm$^{-2}$.  A new optical spectrum is
	     presented, with a \ion{Mg}{ii} absorption doublet detected at
	     $z=0.351$, but no other significant features.  The optical
	     spectrum shows little reddening, implying a low dust to gas
	     ratio in the absorber. MS\,0205.7+3509 must therefore be viewed
	     through a high column density, low-metallicity gas cloud,
	     probably at $z=0.351$ and associated with the galaxy that has
	     been shown to be within $\sim2$\arcsec\ of the BL~Lac.
   }

\begin{document}

   \maketitle
\section{Introduction}

Blazars are divided into BL~Lacs and quasars (either flat-spectrum
radio-loud, optically violently variable, highly-polarised or
core-dominated) based on the strength of emission lines in the optical
spectrum \citep[and references therein]{1997A&A...325..109S}.

It has been suggested that sources with BL Lac characteristics are actually
gravitationally microlensed quasars
\citep{1986A&A...157..383N,1990Natur.344...45O}. In these cases, it is
expected that stellar mass lenses in a foreground galaxy significantly
amplifies the central QSO continuum source but not the emission from the
line-emitting regions and that variations in the relative source-lens
position could account for the rapid variability observed in many BL~Lacs. 
Sources of this kind should clearly have foreground galaxies, which would
result in an apparent decentering of the AGN from the ``host" and an excess
of absorption in these sources. However, the suggestion that BL~Lacs are
gravitationally microlensed quasars can be discounted for most BL~Lacs
\citep{1992MNRAS.257..404P} and only a few remain as possible or probable
candidates, most notably AO\,0235+164 which appears to have foreground
absorption
\citep{1993ApJ...415..101A,1994ApJ...432..554M,1996ApJ...459..156M} and a
companion AGN \citep{1996AJ....112.2533B}. Other candidates include
PKS\,0537-441 \citep[which shows rapid microvariability, but does not show
evidence for a foreground object in optical imaging or
spectroscopy,][]{1999A&AS..135..477R,2002A&A...392..407P}, and B2\,1308+326
which has characteristics intermediate between BL~Lacs and quasars
\citep{1993ApJ...410...39G,2000A&A...364...43W}, but where high resolution
imaging of the BL~Lac with the HST WFPC2 \citep{1999ApJ...512...88U} was
consistent with a point source.

MS\,0205.7+3509 is another such rare candidate, and while deep imaging has
revealed that the BL~Lac is centred on a host galaxy that is likely an
elliptical and not offset in a spiral host as had been inferred from
previous observations \citep{1995ApJ...454...55S}, a companion
galaxy was also detected very close to the BL~Lac line of sight
\citep{1997A&A...321..374F} which had caused the previous inference of
decentering in a spiral host to be made. X-ray observations with \emph{ROSAT}
\citep{1995ApJ...454...55S} and \emph{ASCA} \citep{1999A&A...345..414W}
showed the existence of absorption well above the Galactic level, indeed at
a level second only among BL~Lacs to PKS\,1413+135
\citep{2002AJ....124.2401P}.  These results, on MS\,0205.7+3509, imply that
the X-ray absorber is in the companion galaxy which is foreground to the
AGN.  It has been suggested that stars in the halo of this companion galaxy
could be responsible for microlensing of the BL~Lac
\citep{1997A&A...321..374F,1999A&A...345..414W}.  

In spite of the relatively good spectral resolution of \emph{ASCA},
the redshift of the absorber could not be constrained from those
observations \citep{1999A&A...345..414W}.  A redshift of $z=0.318$ was
proposed based on the tentative detection of a \ion{Ca}{ii} absorption
system reported in the optical spectrum of this source
\citep{1991ApJ...380...49M} indicating the possible redshift either of the
host galaxy or of a foreground absorber and to date this redshift had been
used as the best available \citep{2000AJ....120.1626R,1995ApJ...454...55S}.

Though somewhat lower in terms of column density, MS\,0205.7+3509 is
six times brighter in X-rays than PKS\,1413+135, making it one of the best
available cases in which to study absorption in the hot phase of the ISM in
a galaxy that is not at low redshift. \emph{XMM-Newton} observations were
performed in an attempt to determine the nature of the X-ray absorber in
MS\,0205.7+3509, in particular in the context of a foreground lensing
galaxy.

Results from these data are presented in this paper. Sect.~2 deals with the
observations and the data reduction procedures; results from the X-ray and
optical data are in Sect.~3.  A discussion of these results and a summary of
our conclusions are given in Sect.~4. Uncertainties given are 90\%
confidence limits unless otherwise stated. A flat universe with
$H_{0}=75$\,km\,s$^{-1}$\,Mpc$^{-1}$ and $\Omega_\Lambda=0.7$ are assumed
throughout.

\section{Observations and data reduction}

MS\,0205.7+3509 was observed by \emph{XMM-Newton}
\citep{2001A&A...365L...1J} for 40\,ks and 20\,ks during orbits 217 and 395
with observations starting on 14 Feb.\ 2001, 07:06:41\,UT and 4 Feb.\ 2002,
19:41:06\,UT respectively.

For the first observation, exposures of 38\,ks and 34\,ks duration were made
with each EPIC-MOS \citep{2001A&A...365L..27T} and with the EPIC-pn
\citep{2001A&A...365L..18S} detector respectively; while for the second
observation, the exposures were 17\,ks (each EPIC-MOS) and 15\,ks (EPIC-pn). 
All observations were performed in Full Frame mode, using the Medium filter
for the MOS cameras and the Thin filter for the pn.  Data were processed and
screened in a standard way with the \emph{XMM-Newton} Science Analysis
System version 5.3.0 (SAS).  Only events corresponding to patterns 0--12
were used for the two MOS cameras, while pattern 0--4 events were selected
from the pn data.

A circular extraction region was defined around the centroid position of the
source, with an aperture of 35\arcsec\ radius.  Data from these spatial
regions were used to extract spectra and lightcurves.  An estimate of the
background was derived by using an aperture of 100\arcsec\ radius at a
source-free position close to the source extraction region.

The complete lightcurves were created by binning the data in 70\,s bins
(giving $\sim150$ counts per bin); two other, different time binnings
yielded similar results.  The background for each instrument was subtracted
from the source lightcurve and the per-instrument lightcurves were added.

The extracted spectra were binned to give a minimum of twenty counts per
bin.  Data in the energy range 0.2--10.0\,keV were used. 
The background-subtracted spectrum from each instrument was fit separately
using the Levenberg-Marquardt algorithm in Xspec v11.0.1 and EPIC response
matrices generated with the SAS.

The EPIC-pn 0.2--10.0\,keV count rate was found to be between 1.2--1.5
counts/s; effects due to photon pile-up are
negligible at this flux level ($\lesssim 0.2\%$ of the total counts).

\begin{figure*}
   \centering
   \begin{minipage}[t]{0.77\textwidth}
     \includegraphics[angle=-90,width=\columnwidth,clip]{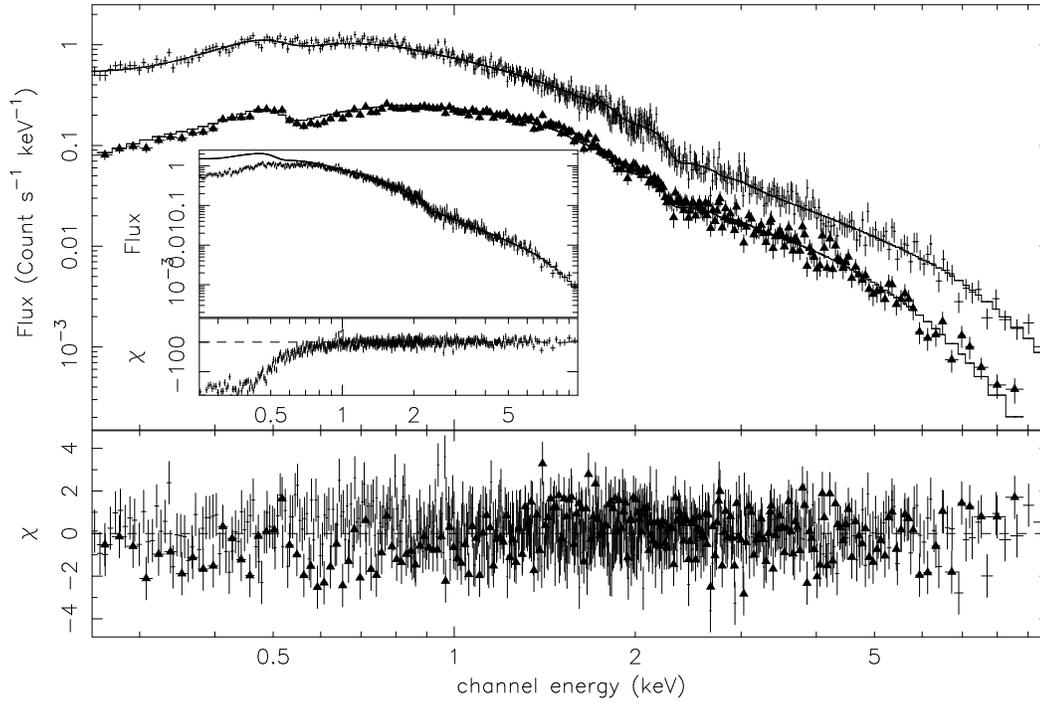}
   \end{minipage}\hfill
   \begin{minipage}[t]{0.22\textwidth}
     \caption{Simultaneous fit of EPIC-pn (crosses) and combined
               EPIC-MOS (triangles) data from 14 Feb.\ 2001 folded through
               the detector response with fit residuals. The model fit was a
               Galactic-absorbed power-law with neutral low-abundance
               absorption at redshift $z=0.351$.  The elemental abundance
               and absorbing column were fit simultaneously in both datasets
               (from 14 Feb.\ 2001 and 4 Feb.\ 2002), whereas the power-law
               slopes and normalisations were allowed to vary independently
               to account for the temporal variability of the synchrotron
               spectrum.  {\em Inset.} Power-law model fit to the 1--10\,keV
               EPIC-pn data of 14 Feb.\ 2001 with Galactic absorption only.
              }
         \label{Fig:sim_fit}
   \end{minipage}
\end{figure*}

Differences between the spectra extracted from the EPIC-MOS cameras are
negligible in this observation.  The MOS data were therefore co-added and
fit as a single spectrum, using the co-added response matrices and
backgrounds.  However, known differences in the cross-calibration of the
EPIC-pn and MOS instruments are apparent in these spectra
(Fig.~\ref{Fig:sim_fit}), these differences primarily affect
the low-energy power-law slopes and normalisations.  In order to exploit the
full dataset, the MOS and pn spectra have been fit separately and then
simultaneously and any significant differences have been highlighted.

\subsection{Optical}
\begin{figure}
   \includegraphics[angle=0,width=\columnwidth,clip]{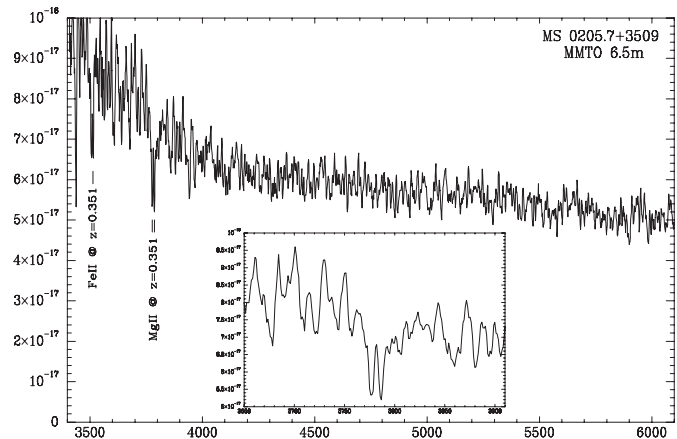}
      \caption{Optical spectrum from the MMTO.  Absorption lines of
               \ion{Fe}{ii} and \ion{Mg}{ii} at $z=0.351$ are
               labelled. \emph{Inset.} A blow-up centred on the resolved
               \ion{Mg}{ii} doublet.  Wavelength (\AA) is plotted on the
               axis of abscissas, flux (arbitrary units) on the ordinate. 
               (The deviations from a power-law shape at the blue and red
               ends of the spectrum are due to calibration and atmospheric
               effects and affect only the continuum shape.)
              }
         \label{Fig:opt_spec}
\end{figure}

An optical spectrum (Fig.~\ref{Fig:opt_spec})
was obtained in a 9200\,s exposure completed on 23 November 2000 with the
Blue Channel spectrograph at the MMTO.\footnote{The MMT Observatory is a
joint facility of the University of Arizona and the Smithsonian Institution}
The slit width was 1\arcsec.25, resulting in a FWHM of 4.2\,\AA. 
The slit position was E-W.  The signal to noise ratio is $\sim20$.  The
spectrum was extracted using the IRAF \textit{kpnoslit} package.  The
spectra were wavelength calibrated with HeNeAr spectra, with a RMS in the
solution of about 0.25\,\AA.  The spectra were flux calibrated with
observations of the standard stars Feige\,34 and G191B2B.

No emission features were identified (W$_\lambda\lesssim3$\AA) in the
spectrum. However, an absorption doublet, identified as \ion{Mg}{ii} at
$z=0.351$, is detected. The first line of the doublet is at 3776.8\,\AA,
with an equivalent width (W$_\lambda$) of $1.3\pm0.1$\,\AA.  The second line
is at 3786.4\,\AA, W$_\lambda = 1.2\pm0.1$\,\AA. The doublet spacing is
correct for the \ion{Mg}{ii} doublet at $z=0.351$ (Fig.~\ref{Fig:opt_spec}
inset).  The individual line widths are not resolved however. A second, low
significance absorption system at 3510\,\AA, corresponding to \ion{Fe}{ii}
2600.17\,\AA\ at $z=0.351$ is also marginally detected
(W$_\lambda=1.6\pm0.2$\,\AA).  We note that the previously reported redshift
($z=0.318$) was tentative and the spectrum very featureless, with lower
signal-to-noise ratio than the spectrum presented here
\citep{1991ApJ...380...49M}.

\section{Results\label{results}}

\subsection{X-ray lightcurves}
The EPIC lightcurve for the first observation is best fit by a linear
increase corresponding to a rise of $\sim 4\%$ during the observation; the
f-test yields a probability of 99.7\% ($3\,\sigma$) for the improvement in
the fit compared to a constant flux value.  The mean 0.2--10\,keV flux for
this observation was $2.80\pm0.01\times10^{-12}$\,erg\,cm$^{-2}$\,s$^{-1}$. 
No significant variability is detected during the second observation, where
the flux was $3.34\pm0.02\times10^{-12}$\,erg\,cm$^{-2}$\,s$^{-1}$, 19\%
higher than the first.  The spectrum hardened between the observations (from
$\Gamma = 2.58\pm0.015$ to $\Gamma = 2.41\pm0.018$), consistent with
previous observations of blazars, where an increase in the synchrotron flux
frequently coincides with an increase in the synchrotron peak frequency
\citep[e.g.][]{1998ApJ...492L..17P,1998ApJ...497..178W}.

\subsection{X-ray spectra}

The spectra from each epoch were found to be consistent, allowing for
variability in the normalisation and slope of the power-laws. Both
observations have therefore been fit simultaneously, allowing only the
power-law component of the model to vary independently. All X-ray spectral
models included neutral absorption fixed at the Galactic level
($6.28\times10^{20}$\,cm$^{-2}$). It is clear however that there is
absorption above the Galactic level in the X-ray spectra of this source
\citep[Fig.~\ref{Fig:sim_fit}, see also][]{1995ApJ...454...55S,1999A&A...345..414W}; fitting a broken power-law
to model the downturn at the soft end of the spectrum instead of absorption
results in a very flat low energy slope ($\Gamma = 0.75$) and a fit
considerably worse than that obtained with neutral absorption at $z=0.351$.

In each observation, fitting a power-law with neutral absorption at
$z=0.351$ to the combined data gives an unacceptable fit statistic ($\chi^2
= 1697$ for 1476 degrees of freedom, null hypothesis probability (NH) =
$5\times10^{-5}$). The data deviate significantly from
the model at low energies, implying that the absorption is different from
that modelled or that the absorber is correct but that there is a different
continuum below $\sim1$\,keV.  Such a difference in continuum at soft
energies is often observed in Seyfert galaxies as a ``soft excess'',
empirically modelled with one or more black-body components
\citep[e.g.][]{2002MNRAS.330L...1P}.  Adding a black-body component to the
power-law with absorption at $z=0.351$ improved the fit significantly
($\chi^2$/DoF$=1584/1474$, NH = 0.02).  The best-fit black-body temperature
was $0.119^{+0.008}_{-0.006}$\,keV with luminosity
$\sim8\times10^{43}$\,erg\,cm$^{-2}$\,s$^{-1}$ at $z=0.351$.  This model
fits the data as well as models with variable redshift or variable
abundances (see Table~\ref{Tab:finalfits}), however the need to invoke an
emission component in order to explain an essentially smooth absorbed
continuum and the fact that a soft excess has never been reported in a
BL~Lac render it a less favoured alternative.

In order to test the nature of the absorber it was assumed that the spectrum
was entirely synchrotron dominated.  The redshift, ionisation state and
metallicity of the absorber were tested in four models: i) a neutral, solar
abundance absorber at $z=0.351$, ii) a neutral, solar abundance absorber at
variable redshift, iii) an ionised, solar abundance absorber at variable
redshift or iv) a neutral, variable abundance absorber at $z=0.351$.  The
results of these four fits to both datasets are presented in
Table~\ref{Tab:finalfits}.
\begin{table}
 \begin{minipage}{\columnwidth}
 \caption{Best-fit parameters and 68\% confidence limits for various fits to 
          the combined EPIC data.  The basic model used was a power-law with
          neutral absorption fixed at the Galactic level
          ($6.28\times10^{20}$\,cm$^{-2}$).  Added to this was either i) a
          neutral, solar abundance absorber at $z=0.351$, ii) a neutral,
          solar abundance absorber at variable redshift, iii) an ionised,
          solar abundance absorber at variable redshift or iv) a neutral
          variable abundance absorber at $z=0.351$.  Columns~1--6 are the
          fitted model, absorbing column density, redshift, abundance and
          ionisation parameter, and the $\chi^2$ statistic over the number
          of degrees of freedom for the fit.}
   \label{Tab:finalfits}
 \setlength{\tabcolsep}{4pt}
 \begin{tabular}{@{}l l c c c r@{}}
 \hline\hline
  	& N$_{\rm H}$			& 		& 	& 	& \underline{$\chi^2$}\\
  Model	& ($10^{20}$\,cm$^{-2}$)	& $z$		& $Z$	& $\xi$	& d.o.f. \\
  \hline
  i)	& 14.5		& 0.351\footnote[6]{Fixed}	& 1.0$^f$	& 0.0$^f$	& \underline{1696.9}\\
	& (14.0--14.9)	& ---				& ---		& ---		& 1476\\[5pt]
  ii)	& 83		& 2.1				& 1.0$^f$	& 0.0$^f$	& \underline{1587.7} \\
	& (70--97)	& (1.90--2.30)			& ---		& ---		& 1475\\[5pt]
  iii)	& 89		& 2.1				& 1.0$^f$	& 0.003		& \underline{1586.1}\\
	& (67--74)	& (1.9--2.2)			& ---		& (0--0.011)	& 1474\\[5pt]
  iv)	& 92		& 0.351$^f$ 			& 0.04		& 0.0$^f$	& \underline{1589.4}\\
	& (81--103)	& ---				& (0.03--0.07)	& ---		& 1475\\
  \hline
  \end{tabular}
 \end{minipage}
\end{table}

\begin{figure}
   \includegraphics[angle=-90,width=\columnwidth,clip]{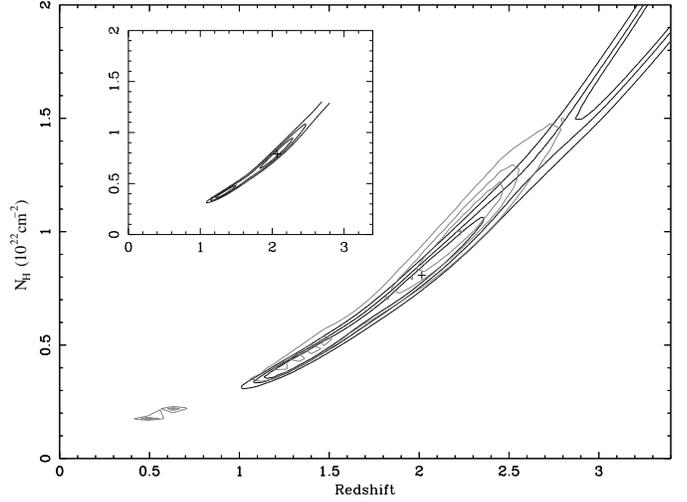}
     \caption{Comparison of MOS-pn differences in the fit parameters,
              redshift and gas absorbing column density.  Confidence
              contours are 1\,$\sigma$ (solid line), 2\,$\sigma$ (dashes)
              and 3\,$\sigma$ (dots) for fits to the EPIC-pn (black) and
              combined EPIC-MOS (grey) spectra from both observations. 
              These parameters are clearly not significantly affected by
              MOS-pn calibration differences. \textit{Inset.} Confidence
              contours for the simultaneous fit of data from all EPIC cameras.
              }
	\label{Fig:z_contours}
\end{figure}

\begin{figure}
   \centering
   \includegraphics[angle=-90,width=\columnwidth,clip]{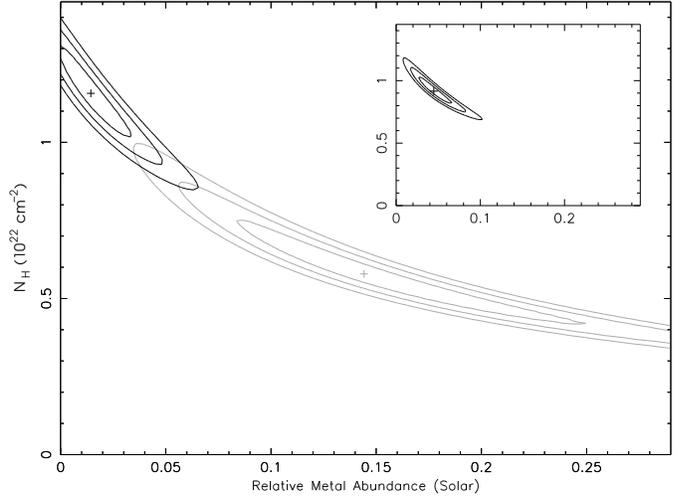}
      \caption{Confidence contours (1, 2 and 3\,$\sigma$) for the best-fit
              metal abundances (relative to the Solar values) and gas
              absorbing column density to the EPIC spectra.  EPIC-pn data
              are plotted in black, MOS in grey.  The contours overlap only
              at greater than the 2\,$\sigma$ level.  However, for both
              datasets the metal abundance is certainly less than half the
              solar value. \textit{Inset.} Confidence contours for the
              simultaneous fit of data from all EPIC cameras.
              }
	 \label{Fig:A_contours}
   \end{figure}

Allowing the redshift to vary (beyond $z = 0.351$) significantly improved
the fit ($\chi^2$/d.o.f. = 1587.7/1475, NH = 0.02, f-test probability
$= 4\times10^{-23}$), giving a best-fit redshift of $z = 2.1\pm0.2$, though
there is another comparable minimum in the $\chi^2$ space near $z=1.35$
(Fig.~\ref{Fig:z_contours}).  The contraint on the redshift arises
principally from the non-detection of a strong neutral oxygen absorption
edge, placing the redshift above one.  In the case where the absorber is
ionised, there is no significant improvement in the fit over the neutral
case ($\chi^2$/d.o.f = 1586.1/1474, NH = 0.02, f-test probability compared
to the neutral absorber = 0.78), and the ionisation parameter is not well
constrained (Table~\ref{Tab:finalfits}).  However, the lack of detection of
strong \ion{O}{vii} or \ion{O}{viii} edges also places the redshift well beyond
$z=0.351$ for this model.

Since it is the non-detection of these edges that constrains the redshift of
the absorber in the latter cases, it is not surprising that a low abundance
absorber with a redshift fixed at $z=0.351$ is statistically as good a fit
as the higher-redshift solar abundance absorbers ($\chi^2$/d.o.f =
1589.4/1475, NH = 0.02). Allowing the redshift to vary in this case did not
significantly improve the fit ($\chi^2$/d.o.f = 1588.9/1474, f-test
probability of 0.50, compared to the model with redshift fixed at
$z=0.351$).  There is some difference in the best-fit metal abundances
determined from the MOS and pn datasets (see Fig.~\ref{Fig:A_contours}) due
to calibration differences between the instruments; in particular, small
systematic residuals may be fit as absorption edges in the MOS spectra.  In
any case, the abundance of the absorber must certainly be below half the
solar value and is probably much lower (at the redshift inferred from the
optical spectrum, $z=0.351$).

\section{Discussion and conclusions}

Observations with \emph{ASCA} indicated the peak of the synchrotron
component in MS\,0205.7+3509 to be between the UV and soft X-rays
at the time of those observations \citep{1999A&A...345..414W}, consistent
with results from the first \emph{XMM-Newton} observation.  The harder
spectrum and higher flux during the second observation imply a shift in the
synchrotron peak frequency to higher energies.

The featureless nature of the absorbed spectrum implies either that the
absorber is metal poor or that the absorbing material is at a redshift much
higher than the $z=0.351$ \ion{Mg}{ii} absorption system detected in the
optical spectrum. In either case the absorbing gas column density is
similarly high ($\sim10^{22}$\,cm$^{-2}$), implying an optical extinction
that is not observed \citep[A$_{\rm V} \simeq 5$ at $z=0$ where the gas to
dust ratio is similar to the Galactic value,][]{1978ApJ...224..132B}, a fact
that supports the low-redshift, low-metallicity model for the absorber.

It is possible that the absorber is associated with the BL~Lac host galaxy. 
But the lack of gas and dust in elliptical galaxies
\citep{1999sfet.conf..119K}, the typical hosts of BL~Lacs, and the proximity
of the companion galaxy \citep[2.3\arcsec,\ corresponding to an apparent
linear distance of $\sim11$\,kpc,][]{1997A&A...321..374F} strongly suggest a
link between the companion and the absorbing gas, as proposed by
\citet{1997A&A...321..374F} and \citet{1999A&A...345..414W}.  Furthermore,
this is, as far as we are aware, the highest column density so far observed
in the spectrum of a BL~Lac object \citep[after PKS\,1413+135 which appears to
have large absorption in its host galaxy, but the host in that case is an
edge-on spiral,][]{2002AJ....124.2401P}, implying that the absorption seen here is not associated with
the BL~Lac host.  In this case the $z=0.351$ \ion{Mg}{ii} belongs to the
X-ray absorbing gas which must be poor in dust and metals, implying that the
BL~Lac is illuminating a fairly pristine gas cloud that is associated with
the galaxy close to the line of sight at redshift $z=0.351$.


\begin{thebibliography}{}

\bibitem[\protect\astroncite{{Abraham} et~al.}{1993}]{1993ApJ...415..101A}
{Abraham} R.~G., {Crawford} C.~S., {Merrifield} M.~R., {Hutchings} J.~B.,
  {McHardy} I.~M., 1993,
\newblock {ApJ} {415}, 101

\bibitem[\protect\astroncite{{Bohlin} et~al.}{1978}]{1978ApJ...224..132B}
{Bohlin} R.~C., {Savage} B.~D., {Drake} J.~F., 1978,
\newblock {ApJ} {224}, 132

\bibitem[\protect\astroncite{{Burbidge} et~al.}{1996}]{1996AJ....112.2533B}
{Burbidge} E.~M., {Beaver} E.~A., {Cohen} R.~D., {Junkkarinen} V.~T., {Lyons}
  R.~W., 1996,
\newblock {AJ} {112}, 2533

\bibitem[\protect\astroncite{{Falomo} et~al.}{1997}]{1997A&A...321..374F}
{Falomo} R., {Kotilainen} J., {Pursimo} T. et~al., 1997,
\newblock {A\&A} {321}, 374

\bibitem[\protect\astroncite{Gabuzda et~al.}{1993}]{1993ApJ...410...39G}
Gabuzda D.~C., Kollgaard R.~I., Roberts D.~H., Wardle J. F.~C., 1993,
\newblock {ApJ} {410}, 39

\bibitem[\protect\astroncite{{Jansen} et~al.}{2001}]{2001A&A...365L...1J}
{Jansen} F., {Lumb} D., {Altieri} B. et~al., 2001,
\newblock {A\&A} {365}, L1

\bibitem[\protect\astroncite{{Knapp}}{1999}]{1999sfet.conf..119K}
{Knapp} G.~R., 1999,
\newblock In: {ASP Conf. Ser. 163: Star Formation in Early Type Galaxies}, pp
  119--134

\bibitem[\protect\astroncite{{Madejski}}{1994}]{1994ApJ...432..554M}
{Madejski} G., 1994,
\newblock {ApJ} {432}, 554

\bibitem[\protect\astroncite{{Madejski} et~al.}{1996}]{1996ApJ...459..156M}
{Madejski} G., {Takahashi} T., {Tashiro} M. et~al., 1996,
\newblock {ApJ} {459}, 156

\bibitem[\protect\astroncite{{Morris} et~al.}{1991}]{1991ApJ...380...49M}
{Morris} S.~L., {Stocke} J.~T., {Gioia} I.~M. et~al., 1991,
\newblock {ApJ} {380}, 49

\bibitem[\protect\astroncite{{Nottale}}{1986}]{1986A&A...157..383N}
{Nottale} L., 1986,
\newblock {A\&A} {157}, 383

\bibitem[\protect\astroncite{{Ostriker} \&
  {Vietri}}{1990}]{1990Natur.344...45O}
{Ostriker} J.~P., {Vietri} M., 1990,
\newblock {Nat} {344}, 45

\bibitem[\protect\astroncite{{Padovani}}{1992}]{1992MNRAS.257..404P}
{Padovani} P., 1992,
\newblock {MNRAS} {257}, 404

\bibitem[\protect\astroncite{{Page} et~al.}{2002}]{2002MNRAS.330L...1P}
{Page} K.~L., {Pounds} K.~A., {Reeves} J.~N., {O'Brien} P.~T., 2002,
\newblock {MNRAS} {330}, L1

\bibitem[\protect\astroncite{{Perlman} et~al.}{2002}]{2002AJ....124.2401P}
{Perlman} E.~S., {Stocke} J.~T., {Carilli} C.~L. et~al., 2002,
\newblock {AJ} {124}, 2401

\bibitem[\protect\astroncite{{Pian} et~al.}{2002}]{2002A&A...392..407P}
{Pian} E., {Falomo} R., {Hartman} R.~C. et~al., 2002,
\newblock {A\&A} {392}, 407

\bibitem[\protect\astroncite{{Pian} et~al.}{1998}]{1998ApJ...492L..17P}
{Pian} E., {Vacanti} G., {Tagliaferri} G. et~al., 1998,
\newblock {ApJ} {492}, L17

\bibitem[\protect\astroncite{{Rector} et~al.}{2000}]{2000AJ....120.1626R}
{Rector} T.~A., {Stocke} J.~T., {Perlman} E.~S., {Morris} S.~L., {Gioia} I.~M.,
  2000,
\newblock {AJ} {120}, 1626

\bibitem[\protect\astroncite{{Romero} et~al.}{1999}]{1999A&AS..135..477R}
{Romero} G.~E., {Cellone} S.~A., {Combi} J.~A., 1999,
\newblock {A\&AS} {135}, 477

\bibitem[\protect\astroncite{{Scarpa} \& {Falomo}}{1997}]{1997A&A...325..109S}
{Scarpa} R., {Falomo} R., 1997,
\newblock {A\&A} {325}, 109

\bibitem[\protect\astroncite{{Stocke} et~al.}{1995}]{1995ApJ...454...55S}
{Stocke} J.~T., {Wurtz} R.~E., {Perlman} E.~S., 1995,
\newblock {ApJ} {454}, 55

\bibitem[\protect\astroncite{{Str{\"u}der} et~al.}{2001}]{2001A&A...365L..18S}
{Str{\"u}der} L., {Briel} U., {Dennerl} K. et~al., 2001,
\newblock {A\&A} {365}, L18

\bibitem[\protect\astroncite{{Turner} et~al.}{2001}]{2001A&A...365L..27T}
{Turner} M. J.~L., {Abbey} A., {Arnaud} M. et~al., 2001,
\newblock {A\&A} {365}, L27

\bibitem[\protect\astroncite{{Urry} et~al.}{1999}]{1999ApJ...512...88U}
{Urry} C.~M., {Falomo} R., {Scarpa} R. et~al., 1999,
\newblock {ApJ} {512}, 88

\bibitem[\protect\astroncite{{Watson} et~al.}{1999}]{1999A&A...345..414W}
{Watson} D., {Hanlon} L., {McBreen} B. et~al., 1999,
\newblock {A\&A} {345}, 414

\bibitem[\protect\astroncite{{Watson} et~al.}{2000}]{2000A&A...364...43W}
{Watson} D., {Smith} N., {Hanlon} L. et~al., 2000,
\newblock {A\&A} {364}, 43

\bibitem[\protect\astroncite{{Wehrle} et~al.}{1998}]{1998ApJ...497..178W}
{Wehrle} A.~E., {Pian} E., {Urry} C.~M. et~al., 1998,
\newblock {ApJ} {497}, 178

\end{thebibliography}
\end{document}